\documentclass[12pt,a4paper]{article}
\usepackage{amsmath}
\usepackage{amsthm}
\usepackage{amssymb}
\usepackage{amsbsy}
\usepackage{amsfonts}
\usepackage{amstext}
\usepackage{amscd}
\usepackage[dvips]{epsfig}

\numberwithin{equation}{section}
\theoremstyle{plain}
\newtheorem{thm}{Theorem}[section]
\newtheorem{prop}[thm]{Proposition}

\theoremstyle{definition}

\newtheorem{rem}[thm]{Remark}

\begin{document}
\title{White noise analysis on manifolds and the energy representation of a gauge group}
\author{Takahiro Hasebe \\ Graduate School of Sciences, Kyoto University \\ Kyoto 606-8502, Japan \\E-mail: hsb@kurims.kyoto-u.ac.jp}
\date{}

\maketitle 

\begin{abstract}
The energy representation of a gauge group on a Riemannian manifold has been discussed by several authors. Y. Shimada has shown the irreducibility for compact Riemannian manifold, using white noise analysis. In this paper we extend its technique to
the noncompact Riemannian manifolds which have differential operators satisfying some
conditions.      
\end{abstract}



\section{Introduction}
The energy representation of a gauge group on a Riemannian manifold has been discussed by 
several authors, for instance, in Refs. 1, 2, 4, 5, 7 and 13. 
 Their methods are essentially to reduce the problem to the estimate of the support of a Gaussian 
measure in an infinite dimensional space. The best result in this direction seems
to be the one in Ref. 13. 
After all these contributions, the irreducibility in two dimensions has remained unsettled yet.
One of the difficulties is the conformal invariance of the energy representation, i.e.,
when we transform the Riemannian metric $g(x)$ into $e^{\rho (x)}g(x)$, the energy representation 
remains unchanged. For a historical survey of this line of research, we refer the 
reader to Ref. 1.

In contrast to the above, Y. Shimada has recently shown the irreducibility 
of a gauge group on a compact Riemannian manifold in Ref. 12, applying white noise analysis. 
Unfortunately, there is a mistake in the proof of Lemma 4.6 in Ref. 12. 
Shimada has used the relation $d\Phi_{s,t} = \beta (\exp(\Phi_{s,t}))$ in the proof of the equation (4.28), 
but this relation does not hold when the Lie group $G$ is non-abelian.  
 
The author could not find a way to overcome this mistake. However, Shimada's approach is still of
importance when we want an analysis of white noise indexed by manifolds, or an analysis of 
the energy representation. Hence we extend this technique to include some 
class of noncompact manifolds in the presence of a weight function. We prove that the energy representation has the differential
representation if a manifold has a differential operator with some
properties. 

\section{Preliminaries}
\subsection{Notation} 
In this section, we explain the notation frequently used throughout this paper.

\begin{itemize}
     \item   $(M,g)$ denotes a Riemannian manifold $M$ equipped with a Riemannian metric $g$.
     \item   $\nabla $ denotes the Levi-Civita connection on $(M,g)$.
     \item   $dv = \sqrt{|g|}dx$ is the Riemannian measure on $(M, g)$. 
     \item   $G, \mathfrak{g}$ denote a compact, semisimple Lie group and its Lie algebra, respectively. 
     \item   $B(\cdot,\cdot)$ means the Killing form of $\mathfrak{g}$. 
     \item   $\mathbb{N}:= \{0, 1, 2, 3,  \cdots \}$.
     \item   $\Gamma_c (T^* M) :=$ the set of all smooth sections of the cotangent bundle on $M$ with supports compact.
     \item   $C^{\infty}_c (M) :=$ the set of all smooth real-valued functions with supports compact.   
     \item   $\langle \cdot,\cdot \rangle _x$ represents the natural bilinear form induced by $g_x$ or $g_x \otimes (-B)$ on tensor products of tangent and cotangent spaces at $x$, and Lie algebra $\mathfrak{g}$, depending on the context. When the complexification $\mathfrak{g}_\mathbb{C}$ is considered, $\langle \cdot,\cdot \rangle _x$ is the natural inner product which is antilinear in the left and        linear in the right. 
     \item   $\langle \cdot , \cdot \rangle _0$ is an inner product on $\Gamma_c(T^{*} M)$ or 
             $\Gamma_c(T^{*} M) \otimes \mathfrak{g}_\mathbb{C}$  determined by 
             $\langle f, g \rangle _0 := \int _M \langle f , g \rangle _x dv(x)$.
     \item   for each $n \geq 1$, $\nabla ^{*}: \Gamma_c({T^{*} M^{\otimes n}}) \longrightarrow  
              \Gamma_c({T^{*} M^{\otimes n-1}})$ 
             is the adjoint operator of  $\nabla$ with respect to the inner product  $\langle \cdot,\cdot \rangle _0$.    
     \item   for each $n \geq 0$, $\Delta  := - \nabla ^{*} \nabla $ is called the 
             Bochner Laplacian on $\Gamma_c({T^{*} M^{\otimes n}})$. 
     \item   $|\omega |_x := \langle \omega, \omega  \rangle _x ^{ \frac{1}{2} }$.
     \item   $C^{\infty}_b (M) := \{h \in C^{\infty} (M) ; \sup_{x \in M} |(\nabla ^m h)(x)|_x <  \infty$ for all $m \in \mathbb{N} \}$. 
     \item   $\Gamma_{\textbf{b}} (X)$ denotes the boson Fock space on $X$, where $X$ is a Hilbert space.     
     \item   $\mathcal{L}(F_1, F_2)$ is the set of all continuous linear operators from a topological 
             vector space $F_1$ to a topological vector space $F_2$. 
\end{itemize}
\subsection{White noise analysis}

We explain white noise analysis needed in this paper.
Let $X$ be a complex Hilbert space equipped with an inner product $\langle \cdot,\cdot \rangle _0$ and $H$ be a self-adjoint 
operator defined on a dense domain $D(H)$ in $X$. Assume that $H$ has $\{\lambda _j \}_{j=1} ^{\infty}$ as 
eigenvalues, and $\{e_j \}_{j=1} ^{\infty}$ as corresponding eigenvectors. 

\vspace{10pt}
\noindent
$\mathbf{Hypothesis}$
\begin{itemize}
    \item[$\cdot$]   $\{e_j \}_{j=1} ^{\infty}$ is a CONS of $X$ 
    \item[$\cdot$]   $ 1 <  \lambda _1 \leq \lambda_2 \leq \cdots \nearrow  \infty$     
\end{itemize} 

Then we can construct a nuclear countably Hilbert space as follows (for details,
the reader is referred to  Ref. 9). 
For $p \in \mathbb{R}$, we can define an inner product $\langle x, y \rangle _p := 
\langle H^p x, H^p y  \rangle _0$ on $D(H^p)$. 
Then $D(H^p)$ becomes a Hilbert space, which we write as $E_p$.
Let $E := \cap_{p \geq 0} E_p$ be a nuclear countably Hilbert space 
equipped with the projective limit topology and let $E^*$ be its dual 
with the strong dual topology. Thus we obtain a Gelfand triple $E \subset X \subset E^*$. 
In the same way, we construct a Gelfand triple $(E) \subset \Gamma_\textbf{b} (X)  \subset (E)^* $
in terms of the self-adjoint operator $\Gamma_\textbf{b} (H)$.

\subsection{The energy representation of a gauge group}
First we define an inner product $\langle f,g\rangle _{\rho
,0}:=\int_{M}\langle f,g\rangle _{x}e^{\rho (x)}dv$ with $\rho \in C^{\infty
}(M)$ for $f,g\in \Gamma _{c}(T^{\ast }M)$ or $\Gamma _{c}(T^{\ast
}M)\otimes \mathfrak{g}_{\mathbb{C}}$. We simply write $\langle f,g\rangle _0$ for 
$\rho = 0$ in accordance with the notation in Section 2.

Let $\mathcal{H}(M;\mathfrak{g}_{ \mathbb{C} })_{\rho}$ be the completion of the space 
$\Gamma _c (T^* M) \otimes \mathfrak{g}_{ \mathbb{C}}$ with the inner product 
$\langle \cdot,\cdot \rangle _{\rho, 0}$. This space is physically the one-particle state space.

For $\psi \in C^{\infty}_c (M ; G)$, the right logarithmic derivative $\beta (\psi)\in 
\Gamma _c (T^* M) \otimes \mathfrak{g}_{ \mathbb{C}}$ is defined as
\begin{gather}
     (\beta (\psi))(x) := d\psi_x \psi(x)^{-1} = R_{\psi(x)^{-1}}d\psi_x.   \\ 
\intertext{$\beta$ satisfies}
     \beta (\psi \phi) = V(\psi)\beta (\phi) + \beta (\psi).
\end{gather} 
The latter equality is said to be the Maurer-Cartan cocycle condition.

For $\psi \in C^{\infty}_c (M ; G)$ and $f \in \mathcal{H}(M;\mathfrak{g}_{ \mathbb{C} })_{\rho}$, let
\begin{equation}
     (V(\psi)f)(x) := [\mathrm{id}_{T^* _x M} \otimes \mathrm{Ad}(\psi(x))]f(x), ~~x \in M,
\end{equation}
then $V$ is a unitary representation of the gauge group $C^{\infty}_c (M ; G)$  on the Hilbert
space $\mathcal{H}(M;\mathfrak{g}_{ \mathbb{C} })_{\rho}$.

Let $U$ be a unitary representation of the gauge group on the boson Fock space 
$\Gamma_\textbf{b} (\mathcal{H}(M;\mathfrak{g}_{ \mathbb{C} })_{\rho})$ determined by 
\begin{equation}
      U(\psi) \exp(f) :=   \exp\Big{(}-\frac{1}{2} |\beta (\psi)|_{\rho, 0} ^2 - \langle \beta (\psi), V(\psi)f \rangle _{\rho, 0}\Big{)}
                             \exp(V(\psi)f + \beta (\psi))
\end{equation}  
for $f \in \mathcal{H}(M;\mathfrak{g}_{ \mathbb{C} })_{\rho}$ and $\psi \in C^{\infty}_c (M ; G)$.
We call this representation the (weighted) energy representation or, if we emphasize the weight 
function, the energy representation  with the weight function $\rho$.

It is important that this representation is, as easily checked,  not a projective representation 
since the Maurer-Cartan cocycle $\beta$ is real.

\begin{rem}
As we stated in Introduction, the energy representation is conformally invariant in 
two dimensions. This is understood as follows.
Let $M$ be a d-dimensional Riemannian manifold. If the Riemannian metric $g$ is 
transformed into $e^{\rho}g$, $dv$ and $\langle \cdot,\cdot \rangle _x$ on $T^* _x M$ are transformed correspondingly: 
\begin{gather}
dv \longrightarrow  e^{ \frac{d}{2} \rho}dv  \\
\langle \cdot,\cdot \rangle _x \longrightarrow e^{- \rho (x)}\langle \cdot,\cdot \rangle _x.          
\end{gather} 
Hence, the inner product $\langle \cdot,\cdot \rangle _{\rho, 0}$ remains invariant if and only if $d = 2$.
Because of the existence of this conformal invariance in two dimensions, the proof of 
irreducibility is difficult. The details are in Refs. 2 and 13.
\end{rem}

\section{Several conditions for a self-adjoint operator}
 
In the following, we show several conditions in order to use white noise analysis on a Riemannian manifold. For this purpose we introduce a function $W$ which tends to infinity in infinite distances. This function and approximately constant functions make the manifold behave as if it is compact. Here the phrase "as if it is compact" means that we can use constant functions, which will be shown in propositions 2 and 3.

Let $M$  be a Riemannian manifold equipped with a Riemannian metric $g$ and $W$ be a positive smooth function. Let $L^2(T^* M)$ denote the completion of the space $\Gamma_c (T^* M)$  with respect to the norm induced by $g$.Note that the quadratic form $Q(f,f) =  \int _M ( |\nabla f |_x^2 + W|f|_x^2) dv(x)$ with domain($Q$) $= \{f: Q(f,f) < \infty \}$ is a nonnegative, symmetric closed form. Hence there is a self-adjoint operator denoted as $H = -\Delta + W$ such that $Q(f,g) = \langle Hf,g \rangle _0 $. First we consider the following condition on $(M,g)$ and $W$. 
\begin{itemize}
   \item[(a)]  $W \in C^{\infty} (M),   ~~W \geq 1 $;
               
              the spectrum of $H = -\Delta + W$ is discrete ({denoted as $\{\lambda _n \}_{n=1} ^\infty$})  
               and satisfies $1 <  \lambda _1 \leq \lambda _2 \leq \cdots \leq \lambda _n \leq \cdots 
               \nearrow \infty$);
               
               there exists $p \geq$ 0 such that $H^{-p}$ belongs to the Hilbert-Schmidt  class.
\end{itemize} 
The condition (a) suffices for compact manifolds. Using this, we can introduce the 
family of seminorms  $\{|\cdot | _p \}_{p \geq 0}$ defined on $\Gamma _c (T^* M)$
(see section 2).
In order to deal with noncompact manifolds, however, (a) is not sufficient. 
Below we introduce a few more conditions. 
Here we define a family of 
seminorms $\{| \cdot |^{\prime} _m \}_{m \in \mathbb{N}}$  defined by
 $|f|^{\prime} _m = \sum_{n=0}^m |W^m \nabla ^n f|_0,  ~~f \in \Gamma _c (T^* M)$.

\begin{itemize}
   \item[(b)]  the two families of seminorms $\{|\cdot | _p \}_{p \geq 0}$ and 
               $\{| \cdot |^{\prime} _{m} \}_{m \in \mathbb{N}}$ 
               define the same topology on $E$;
   \item[(c)]  there exists a sequence $\{\psi _n\}_{n = 1} ^{\infty}$of smooth functions 
               with supports compact, which enjoys the following properties: 
      \begin{itemize}
                \item[$\cdot$] $\psi _n (x) \longrightarrow$ 1 ~~as~~ $n \longrightarrow \infty$,~~ for all $x \in M$,     
                \item[$\cdot$] for every $m \in \mathbb{N}$ there exists $C = C(m)$ independent of $n$ such that 
            \begin{center}    
                     $ \sup_{x \in M} |\nabla ^m \psi _n (x) |_{x} \leq C(m)$ ~~ for all $n \geq 1$.    
            \end{center}       
       \end{itemize}
\end{itemize}  

(b) implies that the space 
$\{f \in L^2(T^* M) ; W^n \nabla ^m f  \in  L^2(\otimes ^m  T^* M) $ for all $ n,m \in \mathbb{N} \}$ 
coincides with the space $E$. (b) is true if the Riemannian manifold $M$ is $\mathbb{R}^d$ or 
compact, with the function $W$ taken as $ |x|^2 + 1 $ and $2$ respectively. Here 
$\Delta $ means the Bochner Laplacian $-\nabla ^* \nabla $. For a proof in the compact case, 
we refer the reader to Ref. 11. 
The Euclidean case is well known. However, for convenience and 
in order to understand the reason why the condition (b) is nontrivial 
in a general Riemannian manifold, we prove this fact for the Euclidean case. 
This fact for the one-dimensional case can be found in Ref. 10 
without a proof.

Let $A_j := \frac{1}{\sqrt{2}}\Big{(} x_j + \frac{\partial}{\partial x_j}\Big{)}$,
    $A_j ^* := \frac{1}{\sqrt{2}}\Big{(} x_j - \frac{\partial}{\partial x_j}\Big{)}$,
    $N_j := A_j ^* A_j $
    and $N := \sum_{j=1} ^d N_j $.
    It holds that $[A_j, A_k ^*] = \delta_{jk}$ and $-\Delta +|x|^2 + 1 = 2N +d+ 1$.
Let $A_j ^{\sharp}$ denote either $ A_j$ or $A_j ^*$, and let $W = |x|^2 + 1$.
We show that there is some $C = C(m)  >  0 $ such that
for $ f \in C_c ^{\infty} (\mathbb{R} ^d)$ and $j_1, \cdots, j_m \in \{1,2,\cdots ,d\} $,
\begin{equation}
      |A_{j_1} ^{\sharp} \cdots A_{j_m} ^{\sharp} f|_0  \leq C(m)|(2N+d+1)^{\frac{m}{2}}f|_0. 
\end{equation}
The proof of (7) results from the canonical commutation relations.
Once (7) is proved, the relations $x_j = \frac{A_j + A_j ^*}{\sqrt{2}}$ and 
$\frac{\partial}{\partial x_j} = \frac{A_j - A_j ^*}{\sqrt{2}}$ lead to the validity of condition (b).

The above argument depends on the properties special to the number operator and creation, annihilation 
operators on $\mathbb{R} ^d$. On a general Riemannian manifold, we do not know how to verify (b) 
(under some mild condition on the manifold),
even if the function $W$ is found to satisfy the condition (a).  

\begin{proof}[Proof of (7)]
We show (7) from an example. 
\begin{gather}
\begin{split}
|A_{1} A_1 ^*  A_2 A_2 f|_0 ^2 &=~ \langle A_1 A_1 ^* A_1 A_1 ^* A_2 ^* A_2 ^* A_2 A_2 f, f \rangle _0   \\
                               &=~ \langle A_1 ^* A_1 A_1 A_1 ^* A_2 ^* A_2 ^* A_2 A_2 f, f \rangle _0 + \langle  A_1 A_1 ^* A_2 ^* A_2 ^* A_2 A_2 f, f \rangle _0     \\
                               &=~ \langle A_1 ^* A_1 A_1 ^* A_1  A_2 ^* A_2 ^* A_2 A_2 f, f \rangle _0 
                                     + \langle A_1 ^* A_1 A_2 ^* A_2 ^* A_2 A_2 f, f \rangle _0        \\
                               &~~~~~~~~~ + \langle  A_1^* A_1  A_2 ^* A_2 ^* A_2 A_2 f, f \rangle _0 + \langle  A_2 ^* A_2 ^* A_2 A_2 f, f \rangle _0         \\          
                               &=~ \langle A_1 ^* A_1 A_1 ^* A_1  A_2 ^* A_2 A_2 ^* A_2 f, f \rangle _0 - \langle A_1 ^* A_1 A_1 ^* A_1  A_2 ^* A_2 f, f \rangle _0   \\
                               &~~~~~~~~~  + 2\langle A_1 ^* A_1 A_2 ^* A_2  A_2 ^* A_2 f, f \rangle _0 - 2\langle A_1 ^* A_1 A_2 ^* A_2 f, f \rangle _0    \\
                               &~~~~~~~~~ +  \langle  A_2 ^* A_2  A_2 ^* A_2 f, f \rangle _0 - \langle  A_2 ^* A_2 f, f \rangle _0                 \\
                               &=~ \langle N_1 ^2 N_2 ^2 f, f \rangle _0 - \langle N_1 ^2 N_2 f, f \rangle _0 + 2\langle N_1 N_2^2 f, f \rangle _0 \\
                               &~~~~~~~~~ - 2\langle N_1 N_2 f, f \rangle _0    +  \langle  N_2 ^2 f, f \rangle _0 - \langle  N_2 f, f \rangle _0                 \\
                               & \leq~ \langle N_1 ^2 N_2 ^2 f, f \rangle _0 + \langle N_1 ^2 N_2 f, f \rangle _0 + 2\langle N_1 N_2^2 f, f \rangle _0 \\
                               &~~~~~~~~~~ + 2\langle N_1 N_2 f, f \rangle _0    +  \langle  N_2 ^2 f, f \rangle _0 + \langle  N_2 f, f \rangle _0                 \\
                               & \leq~ \langle (2N + d + 1)^4 f, f \rangle _0    \\
                               &=~ |(2N+d+1)^2 f|_0 ^2.
\end{split}
\end{gather}
\end{proof}
\begin{rem} If we replace the Schr\"{o}dinger operator with an elliptic operator,  propositions and theorems in this section still hold. However, for simplicity, we restrict ourselves to Schr\"{o}dinger operators.  
\end{rem}








An example of manifolds where the condition (c) fails to hold is $\mathbb{R}^2 \setminus (0, 0)$. 
In fact, if we try to make a sequence 
$\{ {\psi_n} \}_{n = 1} ^{\infty} \subset C^{\infty}_c (\mathbb{R}^2 \setminus (0, 0))$ 
which tends to 1 pointwise, then derivatives of $\psi_n $ tend to infinity near $(0, 0)$ as $n$ 
tends to infinity. \\

Next we consider the weighted representation case.
Put $\nabla _\rho:=  e^{- \frac{\rho}{2} } \circ \nabla \circ e^{ \frac{\rho}{2} } $, 
then we have $\nabla _\rho ^* =  e^{- \frac{\rho}{2} } 
 \circ \nabla ^* \circ e^{ \frac{\rho}{2} } $, in fact for all $f,g \in \Gamma_c (T^* M)$,
\begin{equation}
\begin{split}
         \langle f, \nabla _\rho ^{*}g \rangle _{\rho , 0}~~ &= \langle \nabla _\rho f, g \rangle _{\rho , 0}~~    \\
                                               &= \int_M \langle \nabla _\rho  f,g \rangle _x e^{\rho (x)}dv      \\
                                               &= \int_M \langle  e^{- \frac{\rho}{2} } \nabla (e^{ \frac{\rho}{2} } f),g \rangle _x e^{\rho (x)}dv       \\
                                               &= \int_M \langle \nabla (e^{ \frac{\rho}{2} } f), e^{ \frac{\rho}{2} } g \rangle _x dv   \\    
                                               &= \int_M \langle  (e^{ \frac{\rho}{2} } f), \nabla ^* (e^{ \frac{\rho}{2} } g) \rangle _x dv       \\
                                               &= ~~~\langle f, e^{- \frac{\rho}{2} } \nabla ^* ( e^{ \frac{\rho}{2}}g)  \rangle _{\rho, 0},   \\
\end{split}
\end{equation}
where $*$ on the left and right hand sides mean adjoints in  $\Gamma _c (T^* M)_{\rho}$ and 
$\Gamma_c(T^* M)$ respectively. Let  $H_{\rho}:= \nabla _\rho ^* \nabla _\rho + W$ and  $e_{\rho, n} := e^{- \frac{\rho}{2} }e_n $
, then we have  $H_{\rho} e_{\rho, n} = \lambda_n e_{\rho, n}$. 
Correspondingly, $|\cdot|_p$ and $|\cdot|^{\prime}_m $ are replaced with $|f|_{\rho, p} := |H_{\rho}^p f|_{\rho,0}$ and
$|f|^{\prime} _{\rho, m} := \sum_{n=0}^m |W^m \nabla _\rho ^n f|_{\rho, 0}$ respectively. 
With the above replacements, we can prove the following properties easily:
\begin{itemize}
   \item[(a$^\prime$)] the spectrum of $H_{\rho} = \nabla _\rho ^* \nabla _\rho + W$ 
              is $\{\lambda _n \}_{n=1} ^\infty$  and                           
               there exists $p \geq$ 0 such that $H_{\rho}^{-p}$ belongs to the Hilbert-Schmidt  class,

   \item[(b$^\prime$)]  the two families of seminorms $\{|\cdot | _{\rho, p}\}_{p \geq 0}$ and 
               $\{| \cdot |^{\prime} _{\rho, m} \}_{m \in \mathbb{N}}$ 
               define the same topology on $E_\rho$, 
\end{itemize}
where $E_\rho$ is defined in the same way as $E$. 
(a$^\prime$) is obvious. 
The remaining (b$^\prime$) is easily checked; for instance,
\begin{equation}
\begin{split}
        |W^m \nabla_\rho ^n f|_{\rho, 0} &= |e^{- \frac{\rho}{2} } W^m \nabla ^n (e^{ \frac{\rho}{2} } f)|_{\rho, 0}    \\
                                                &= |W^m \nabla ^n (e^{ \frac{\rho}{2} } f)|_{0}    \\                   
                                                & \leq  C|e^{ \frac{\rho}{2} } f|_p  ~~~~~~~(\exists C  > 0, \exists p \geq 0)  \\
                                                &= C|H_{\rho}^p f|_{\rho, 0}         \\
                                                &= C|f|_{\rho,p}.        
\end{split}
\end{equation}
 
 \section{The differential representation for an energy representation}
The next theorem is essentially due to Ref. 12, but the proof is changed slightly in the present case. 
This theorem allows us to differentiate the representation of the gauge group.

\begin{thm}  Let $\rho$ be a smooth function on $M$. 
Assume that the conditions (a) and (b) hold. 
Let $\psi _t (x) := \exp (t\Psi (x))$ for $\Psi \in C^{\infty} _c (M ;\mathfrak{g} )$.
Then $ \{V(\psi_t)\}_{t \in \mathbb{R}}$ is a regular one-parameter subgroup of $GL(E_\rho)$, namely, for any 
$p \geq 0$ there exists $q \geq 0$ such that
\begin{gather}
       \lim_{t \rightarrow 0} \sup _{f \in E_\rho , |f|_{\rho, q} \leq 1} \bigg{|}\frac{V(\psi _t) f - f}{t} - V^{\prime}(\Psi) f \bigg{|}_{\rho, p} \longrightarrow 0,  \\
\intertext{where}
        (V^{\prime} (\Psi)f) (x) := [\mathrm{id}_{T^* _x M} \otimes \mathrm{ad}(\Psi(x))] f(x),~~~~   x \in M,  ~~~~ f \in E_\rho .
\end{gather}
\end{thm}
\begin{proof} We only prove the case $\rho = 0$. The proof for nonzero $\rho$ is the same.
From the condition (b), it is sufficient to prove for seminorms $|\cdot|^{\prime}_m, m \in \mathbb{N} $            .
First we show that for $\Psi \in C^{\infty} _c (M ;\mathfrak{g})$ and $m \in \mathbb{N}$, there exists $C = C(\Psi ,m)  >  0$ such that
\begin{gather}
        |V^{\prime}(\Psi)f|^{\prime}_{m} \leq C(\Psi ,m) |f|^{\prime}_m, ~~~~  f \in E_\rho.  \\
\intertext{The above inequality is  proved as follows. For simplicity, we use the same $C$ or $C(\Psi , m)$ in different lines.}
   \begin{split}
           |V^{\prime}(\Psi)f|^{\prime}_m &= \sum_{n=0}^m |W^m \nabla ^n (\Psi f - f \Psi)|_0   \\
                                 &\leq  C(m)  \sum_{n=0}^m \sum_{k=0}^n (|W^m \nabla ^k \Psi \nabla ^{n-k} f|_0 + |W^m \nabla^{n-k} f \nabla ^k \Psi |_0)   \\
                                 &\leq  C(\Psi , m)\sum_{n=0}^m  \sum_{k=0}^n (|W^m \nabla ^{n-k} f|_0 + |W^m \nabla^{n-k} f |_0)   \\
                                 &\leq  C(\Psi , m)|f|^{\prime}_m.
\end{split}
\end{gather}
In particular, $V^{\prime}(\Psi)$ belongs to $\mathcal{L}(E_\rho, E_\rho)$. Hence,
\begin{equation*}
    \begin{split}
        V(\psi_t)f(x) &= [\mathrm{id}_{T^* _x M} \otimes \mathrm{Ad}(\exp(t\Psi(x)))] f(x)  \\
                      &= [\mathrm{id}_{T^* _x M} \otimes \exp(t \mathrm{ad}(\Psi(x)))] f(x)  \\
                      &= \sum_{k=0}^{\infty} \Big{[}\mathrm{id}_{T^* _x M} \otimes \frac{1}{k!}(t \mathrm{ad}(\Psi (x)))^k\Big{]} f(x)  \\   
                      &= \sum_{k=0}^{\infty} \Big{[} \frac{1}{k!}(tV^{\prime}(\Psi))^k f\Big{]}(x),    
\end{split}
\end{equation*}                      
so that 
\begin{equation*}
\begin{split}
       \bigg{|} \frac{V(\psi _t)f - f}{t} - V^{\prime}(\Psi)f \bigg{|}^{\prime} _m &\leq t \sum_{k=2}^{\infty} \frac{1}{k!}t^{k-2} |V^{\prime} (\Psi)^k f|^{\prime} _m   \\                            
                                                                                   &\leq t \sum_{k=2}^{\infty} \frac{1}{k!}t^{k-2} C(\Psi, n)^k |f|^{\prime} _m    \\
                                                                                   &\leq t \exp(C(\Psi, m)) |f|^{\prime} _m.        
\end{split}
\end{equation*}
Then the conclusion of the proposition follows immediately. 
\end{proof}

The next result enables us to use constant functions in $C^{\infty}_b (M;\mathfrak{g})$, as used by Shimada for calculations of commutants of the representation.  This is a consequence of the conditions (b) and (c).

\begin{prop} Let $\rho$ be a smooth function on $M$. 
Let $\Psi$ be an element in $ C^{\infty}_b (M;\mathfrak{g})$. 
The operator  
\begin{equation}
     f \longmapsto  V^{\prime}(\Psi)f 
\end {equation}
belongs to $\mathcal{L}(E_\rho , E_\rho)$ and there exists a sequence $\{\Psi_n \}_{n=1}^{\infty}$ of 
$\mathfrak{g}$-valued smooth functions with supports compact such that
\[
V^{\prime}(\Psi - \Psi _n)f \longrightarrow 0 \text{~in~} E_\rho \text{~as~} n \longrightarrow \infty \text{~for all~} f \in E_\rho.
\]
\end{prop}
\begin{proof}  Again we prove only for $\rho = 0$. Let $\Psi$ be a fixed element in $C^{\infty}_b (M;\mathfrak{g})$ and
$\{\psi_n \}^{\infty} _{n=1}$ be the sequence in the condition (c). We define 
$\Psi_n := \psi_n \Psi \in C^{\infty}_c (M;\mathfrak{g})$.
\begin{equation}
    \begin{split}
        |V^{\prime}(\Psi - \Psi_n)f|_{p} &=  | H^p ((\Psi - \Psi_n)f -f(\Psi - \Psi_n))|_0       \\                
                                &\leq  |H^p ((\Psi - \Psi_n)f)|_0 + |H^p (f(\Psi - \Psi_n))|_0       \\
                                &=   \langle H^{2p} ((\Psi - \Psi_n)f) , (\Psi - \Psi_n)f \rangle _0 ^{\frac{1}{2}}  + \langle H^{2p} (f(\Psi - \Psi_n)), f(\Psi - \Psi_n) \rangle _0 ^{\frac{1}{2}}  \\
                                &\leq  |(\Psi - \Psi_n)f|_{2p} ^{\frac{1}{2}} |(\Psi - \Psi_n)f|_0 ^{\frac{1}{2}} +  |f(\Psi - \Psi_n)|_{2p} ^{\frac{1}{2}} |f(\Psi - \Psi_n)|_0 ^{\frac{1}{2}}.   \\               
\intertext{Schwarz's inequality was used in the last line. We recall that there exist 
$C  > 0$ and $m \in \mathbb{N}$ such that $|h|_{2p} \leq C |h|^{\prime} _m$ for all $h \in E_\rho$ 
and, from the condition (c), that for every $l \in \mathbb{N}$ there exists $C = C(l)$ independent of $n$ such that
\[ \sup_{x \in M} |\nabla ^l (\Psi - \Psi _n) (x)|_{x} \leq C(l) \] for all $n \geq 1$.  Therefore, we have }
        |V^{\prime}(\Psi - \Psi_n)f|_{p} &\leq C |(\Psi - \Psi_n)f|_m ^{\prime \frac{1}{2} } |(\Psi - \Psi_n)f|_0 ^{\frac{1}{2} } +  C |(f(\Psi - \Psi_n))|_m ^{\prime \frac{1}{2} } |f(\Psi - \Psi_n)|_0 ^{\frac{1}{2} }   \\               
                                &= C \bigg{(}\sum_{k=0} ^m |W^m \nabla ^k((\Psi - \Psi_n) f)|_0\bigg{)} ^{ \frac{1}{2} } |(\Psi - \Psi_n)f|_0 ^{\frac{1}{2} }     \\
                                &~~~~~~~~~      +  C \bigg{(}\sum_{k=0} ^m|W^m \nabla ^k(f(\Psi - \Psi_n))|_0 \bigg{)} ^{\frac{1}{2} } |f(\Psi - \Psi_n)|_0 ^{\frac{1}{2} }   \\               
                                &\leq C |(\Psi - \Psi_n)f|_0 ^{\frac{1}{2} } +  C |f(\Psi - \Psi_n)|_0 ^{\frac{1}{2} }.   
\end{split}
\end{equation}
Applying Lebesgue's bounded convergence theorem, we get the desired result. 
\end{proof}

\section*{Acknowledgments}     
The author expresses his sincere thanks to Professor Izumi Ojima for
inspiring discussions about energy representations in terms of gauge
theory. He thanks Mr. Hiroshi Ando, Mr. Hayato Saigo and Mr. Ryo Harada
for useful discussions. He also thanks Professor Takeyuki Hida for
instruction on white noise analysis during his visits to RIMS. This work
was supported by Grant-in-Aid for JSPS research fellows (21-5106).

\paragraph{References}
\begin{itemize}
    \item[1.] S. Albeverio, R. H{\o}egh-Krohn, J. Marion, D. Testard and B. Torr\'{e}sani,
              \textit{Noncommutative Distributions. Unitary Representation of Gauge Groups and Algebras.}
               Marcel Dekker, 1993.
    \item[2.] S. Albeverio, R. H{\o}egh-Krohn and D. Testard, Irreducibility and reducibility for
              the energy representation of the gauge group of mappings of a Riemannian manifold
               into a compact semisimple Lie group, \textit{J. Func. Anal.} \textbf{41} (1981) 378-396.  
    \item[3.] I. Chavel, \textit{Riemannian Geometry. A Modern Introduction, Second Edition.} Cambridge
                University Press, 2006.
    \item[4.] I. M. Gelfand, M. I. Graev and  A. M. Ver\v{s}ik, Representations of the group of smooth mappings
              of a manifold X into a compact Lie group, \textit{Compositio Math.} \textbf{35} (1977) 299-334. 
    \item[5.] I. M. Gelfand, M. I. Graev and  A. M.Ver\v{s}ik, Representations of the group of functions taking 
             values in a compact Lie group, \textit{Compositio Math.} \textbf{42} (1980/1981) 217-243.
    \item[6.] T. Hida, H. -H. Kuo, J. Potthoff, and L. Streit, \textit{White Noise. An infinite dimensional
               calculus.} Kluwer, Dordrecht, 1993.  
    \item[7.] R. S. Ismagilov, Unitary representations of the group $C_0 ^{\infty} (M,G)$, $G =$ SU$_2$,
              \textit{Mat. Sb.} \textbf{100} (1976) 117-131; English transl. in \textit{Math. USSR-Sb.} \textbf{142} (1976) 105-117.
    \item[8.] A. Knapp, \textit{Lie groups beyond an introduction, 2nd ed}. Birkh\"{a}user, 2002. 
    \item[9.] N. Obata, \textit{White Noise Calculus and Fock Space}, Lecture Notes in Math, Vol 1577.
               Springer-Verlag, 1994.
    \item[10.] M. Reed and B. Simon, \textit{  Methods of Modern Mathmatical Physics. vol.1,
               Functional Analysis.} Academic Press, 1972. 
    \item[11.] G. Salomonsen, Equivalence of Sobolev spaces, \textit{Res. Math.} \textbf{39} (2001) 115-130.
    \item[12.] Y. Shimada, On irreducibility of the energy representation of the gauge group
                  and the white noise distribution theory, \textit{Inf. Dim. Anal.
                  Quantum Probab. Rel. Topics.} \textbf{8} (2005) 153-177.  
    \item[13.] N. Wallach, On the irreducibility and inequivalence of unitary representations of 
            gauge groups, \textit{Compositio Math.} \textbf{64} (1987) 3-29.  
\end{itemize}

\end{document}